\documentclass[a4paper,12pt]{article}
\usepackage{amsfonts,amsmath,amssymb,epsf}
\usepackage{amsmath}
\usepackage{graphicx,color}
\usepackage{color}
\begin{document}

\newtheorem{definition}{Definition}[section]
\newcommand{\be}{\begin{equation}}
\newcommand{\ee}{\end{equation}}
\newcommand{\bea}{\begin{eqnarray}}
\newcommand{\eea}{\end{eqnarray}}
\newcommand{\LE}{\left[}
\newcommand{\R}{\right]}
\newcommand{\nn}{\nonumber}
\newcommand{\Tr}{\text{Tr}}
\newcommand{\N}{\mathcal{N}}
\newcommand{\G}{\Gamma}
\newcommand{\vf}{\varphi}
\newcommand{\LL}{\mathcal{L}}
\newcommand{\Op}{\mathcal{O}}
\newcommand{\HH}{\mathcal{H}}
\newcommand{\arctanh}{\text{arctanh}}
\newcommand{\up}{\uparrow}
\newcommand{\down}{\downarrow}
\newcommand{\rd}{\partial}
\newcommand{\de}{\partial}
\newcommand{\ba}{\begin{eqnarray}}
\newcommand{\ea}{\end{eqnarray}}
\newcommand{\db}{\bar{\partial}}
\newcommand{\we}{\wedge}
\newcommand{\ca}{\mathcal}
\newcommand{\lr}{\leftrightarrow}
\newcommand{\f}{\frac}
\newcommand{\s}{\sqrt}
\newcommand{\vp}{\varphi}
\newcommand{\hvp}{\hat{\varphi}}
\newcommand{\tvp}{\tilde{\varphi}}
\newcommand{\tp}{\tilde{\phi}}
\newcommand{\ti}{\tilde}
\newcommand{\ap}{\alpha}
\newcommand{\pr}{\propto}
\newcommand{\mb}{\mathbf}
\newcommand{\ddd}{\cdot\cdot\cdot}
\newcommand{\no}{\nonumber \\}
\newcommand{\la}{\langle}
\newcommand{\lb}{\rangle}
\newcommand{\ep}{\epsilon}
 \def\we{\wedge}
 \def\lr{\leftrightarrow}
 \def\f {\frac}
 \def\ti{\tilde}
 \def\ap{\alpha}
 \def\pr{\propto}
 \def\mb{\mathbf}
 \def\ddd{\cdot\cdot\cdot}
 \def\no{\nonumber \\}
 \def\la{\langle}
 \def\lb{\rangle}
 \def\ep{\epsilon}
\def\m{{\mu}}
 \def\w{{\omega}}
 \def\n{{\nu}}
 \def\ep{{\epsilon}}
 \def\d{{\delta}}
 \def\rh{\rho}
 \def\t{{\theta}}
 \def\a{{\alpha}}
 \def\T{{\Theta}}
 \def\frac#1#2{{#1\over #2}}
 \def\l{{\lambda}}
 \def\G{{\Gamma}}
 \def\D{{\Delta}}
 \def\g{{\gamma}}
 \def\s{\sqrt}
 \def\ch{{\chi}}
 \def\b{{\beta}}
 \def\CA{{\cal A}}
 \def\CC{{\cal C}}
 \def\CI{{\cal I}}
 \def\CO{{\cal O}}
 \def\o{{\rm ord}}
 \def\Ph{{\Phi }}
 \def\L{{\Lambda}}
 \def\CN{{\cal N}}
 \def\p{\partial}
 \def\pslash{\p \llap{/}}
 \def\Dslash{D \llap{/}}
 \def\Mp{m_{{\rm P}}}
 \def\apm{{\alpha'}}
 \def\r{\rightarrow}
 \def\Re{{\rm Re}}
 \def\MG{{\bf MG:}}
\def\be{\begin{equation}}
\def\ee{\end{equation}}
\def\ba{\begin{eqnarray}}
\def\ea{\end{eqnarray}}
\def\bal{\begin{align}}
\def\eal{\end{align}}

 \def\de{\partial}
 \def\db{\bar{\partial}}
 \def\we{\wedge}
 \def\lr{\leftrightarrow}
 \def\f {\frac}
 \def\ti{\tilde}
 \def\ap{\alpha}
  \def\al{\alpha'}
 \def\pr{\propto}
 \def\mb{\mathbf}
 \def\ddd{\cdot\cdot\cdot}
 \def\no{\nonumber \\}
\def\nn{\nonumber \\}
 \def\la{\langle}
 \def\lb{\rangle}
 \def\ep{\epsilon}
 \def\ddbp{\mbox{D}p-\overline{\mbox{D}p}}
 \def\ddbt{\mbox{D}2-\overline{\mbox{D}2}}
 \def\ov{\overline}
 \def\cl{\centerline}
 \def\vp{\varphi}
\def\hB{\hat \Box}

\begin{titlepage}

\thispagestyle{empty}

\begin{flushright}
YITP-16-107\\
IPMU16-0136\\
\end{flushright}

\vspace{.4cm}
\begin{center}
\noindent{\large \textbf{From Path Integrals to Tensor Networks for AdS/CFT}}\\
\vspace{2cm}

Masamichi Miyaji $^{a}$,
Tadashi Takayanagi$^{a,b}$,
Kento Watanabe $^{a}$
\vspace{1cm}

{\it
$^{a}$Center for Gravitational Physics, Yukawa Institute for Theoretical Physics (YITP),\\
Kyoto University, Kyoto 606-8502, Japan\\
$^{b}$Kavli Institute for the Physics and Mathematics of the Universe (Kavli IPMU),\\
University of Tokyo, Kashiwa, Chiba 277-8582, Japan\\
}

\vskip 2em
\end{center}

\vspace{.5cm}
\begin{abstract}
In this paper, we discuss tensor network descriptions of AdS/CFT from two different viewpoints. First, we start with an Euclidean path-integral computation of ground state wave functions with a UV cut off. We consider its efficient optimization by making its UV cut off position dependent and define a quantum state at each length scale. We conjecture that this path-integral corresponds to a time slice of AdS. Next, we derive a flow of quantum states by rewriting the action of Killing vectors of AdS$_3$ in terms of the dual 2d CFT. Both approaches support a correspondence between the hyperbolic time slice H$_2$ in AdS$_3$ and a version of continuous MERA (cMERA). We also give a heuristic argument why we can expect a sub-AdS scale bulk locality for holographic CFTs.
\end{abstract}

\end{titlepage}

\newpage

\section{Introduction}

Even though the holographic principle \cite{Hol}, especially the AdS/CFT correspondence \cite{Ma} is expected to provide us with an extremely powerful method to understand quantum gravity, its basic mechanism has been still mysterious. One interesting possibility of explaining the mechanism of holography is its possible connections to tensor networks. Tensor networks are methods to express quantum wave functions in terms of network diagrams and are in a very similar sprit of holography because it geometrizes algebraically complicated quantum states.

In the pioneering work \cite{Swingle}, it has been conjectured that the AdS/CFT correspondence may be interpreted as the MERA (multi-scale entanglement renormalization ansatz) network \cite{MERA,RMERA}, which is a particular example of tensor networks for conformal field theories (CFTs). At an intuitive level, this fits nicely with the AdS/CFT in that the MERA network is aimed at an explicit real space renormalization group (RG) flow in terms of RG evolution of quantum states. Moreover, estimations of entanglement entropy of MERA networks done in \cite{MERA}, look analogous to the holographic entanglement entropy \cite{RT}, realizing emergent spacetimes from quantum entanglement (for recent reviews see \cite{Ra,RaT}). Such a connection to tensor networks can also be strongly suggested by the recent reformulation of holographic entanglement entropy in terms of bits threads \cite{FH}.

The tensor network is defined in a discretized lattice such as spin systems and to connect the actual AdS/CFT we need to take a continuum limit. A candidate of a continuum version of MERA is formulated in \cite{cMERA} and is called continuous MERA (cMERA). It was conjectured in \cite{NRT,MNSTW} that the AdS/CFT can be regarded as a cMERA network. Even though the special conformal invariance is not realized in the MERA network, cMERA has an advantage that this symmetry is clearly realized.

In the original argument \cite{Swingle}, the MERA network was considered to describe the canonical time slice of AdS$_{d+2}$ i.e. hyperbolic space H$_{d+1}$. However later, it has been pointed out by several authors \cite{Beny,Cz,Car,VP} that the MERA network may correspond to a de Sitter spacetime dS$_{d+1}$ instead of hyperbolic space, especially from the viewpoint of its causal structure of MERA. Note that a hyperbolic space and de Sitter spacetime have the same isometry $SO(1,d+1)$ and it is not easy to distinguish them only by symmetries. In the paper \cite{Cz}, the MERA tensor network is argued to describe a space called kinematical space, which is non-locally related to the original AdS spacetime with mathematically rich structures \cite{Czt}.

Nevertheless, we are still tempting to interpret an AdS spacetime itself as a continuous limit of a certain tensor network (a continuous tensor network, which we call cTN). A quantum state is time-evolved by a given Hamiltonian and this is also described by a network of unitary transformations. Thus we can have a tensor network description of whole spacetime if each time slice, which defines a quantum state, is described by a tensor network. A powerful advantage of tensor network description is that we can take a subregion inside the network and define a quantum state by contracting tensors.
Motivated by this, in \cite{MT}, it was conjectured that in any spacetime described by Einstein gravity, each codimension two convex surface corresponds to a quantum state in the dual theory, called surface/state correspondence. This largely extends the holographic principle as it can be applied to gravitational spacetime without any boundaries. The perfect tensor network \cite{HAPPY} (see also closely related network using random tensors \cite{HQ}), found from a relation between quantum error correcting codes and holography, provides an explicit toy example for the surface/state correspondence. Moreover, it is expected to provide a tensor network which corresponds to the hyperbolic space $H_2$ rather than the de Sitter spacetime. Indeed, it has a discretized version of the full conformal symmetry. Even though this network respects the isometry of AdS$_3$ and holographic entanglement entropy, the quantum state itself has a flat entanglement spectrum and thus deviates from any vacuum of CFTs.

To understand connections between tensor networks and AdS/CFT better, in this paper we would like to
study the AdS$_3/$CFT$_2$ duality from two different viewpoints: (i) Euclidean path-integral description of wave function with a position dependent cut off, and (ii) $SL(2,R)$ transformations of AdS$_3$ in surface/state correspondence. We will see that both approaches support the correspondence between a hyperbolic time slice in AdS$_3$ and a cMERA-like network for a CFT vacuum. The second approach also shows that we can equally identify the network with a de Sitter slice. Next we turn to the excited states. Especially we focus on the locally excited state in the bulk AdS$_3$ and cTN description of its CFT dual. As we will see below, we will find a consistent picture which reveals
 a structure similar to the perfect tensor network \cite{HAPPY} and the random tensor network \cite{HQ}. Finally we will give a heuristic argument why we expect a sub-AdS scale bulk locality for holographic CFTs. In this paper we mainly consider two dimensional CFTs (2d CFTs) for simpler presentations. However, many results can be generalized into the higher dimensional AdS/CFT.

This paper is organized as follows. In section two, we give a brief review of AdS$_3$ geometry and
its global symmetry. In section three, we study an Euclidean path-integral description of vacuum wave function of 2d CFT and introduce a position dependent cut off, which preserves the conformal symmetries. We argue this corresponds to the time slice of AdS space. Following this approach we calculate a wave function at each length scale. In section four, we review the formulation of cMERA with several updates. In section five, we identify a continuous tensor network which describes the global AdS$_3$ spacetime via the Killing symmetry of the AdS space. In section six, we give a heuristic argument why we expect a sub-AdS scale bulk locality for holographic CFTs. In section seven, we examine locally excited states in the bulk in our continuous tensor network description.  In section eight, we summarize our conclusions and discuss future problems. In appendix A, we present details of several choice of cut off functions in Euclidean path-integrals. In appendix B, we extend the construction of cMERA to present a formulation of continuous tensor networks which describe a holographic spacetime and show consistency conditions.

When we were completing this paper, we noticed a very interesting paper \cite{FLP} where a connection between continuous tensor networks and wave functions in Euclidean path-integral with a UV cut off have been studied. The path-integral formulation in our paper is different from theirs in that we made the UV cut off position dependent. The appendix A in this paper includes a consistency
between our work and \cite{FLP}.

\section{AdS$_3$ Geometry}

In this section, we briefly review basic properties of the AdS$_3$ geometry, which will be important in our later arguments. The Lorentzian AdS$_3$ space with a radius $L$ is defined by the hypersurface in $R^{2,2}$:
\be
X^2_0+X^2_3=X^2_1+X^2_2+L^2.  \label{surface}
\ee

The global AdS$_3$ (with radius $L$) is defined by the parametrization
\ba
&& X_0=L\cosh\rho\cos t, \no
&& X_3=L\cosh\rho\sin t,\no
&& X_1=L\sinh\rho\sin\phi, \no
&& X_2=L\sinh\rho\cos\phi.
\ea
This leads to the following metric:
\be
ds^2=L^2(-\cosh^2\rho dt^2+d\rho^2+\sinh^2\rho d\phi^2). \label{gads}
\ee
The Euclidean global AdS$_3$ (i.e. $H_3$) is obtained by the Wick rotation $t\to it$.

Sometimes it is also useful to work with the Poincare AdS metric
\be
ds^2=L^2\f{dz^2-d\tau^2+dx^2}{z^2}, \label{poi}
\ee
in our later arguments.

\subsection{Global Symmetry}

The AdS$_3$ space (\ref{gads}) has the  $SL(2,R)_L\times SL(2,R)_R$ symmetry, which are generated by
 $(L_1,L_0,L_{-1})$ and $(\ti{L}_1,\ti{L}_0,\ti{L}_{-1})$. They correspond to the (global) Virasoro symmetry of its dual 2d CFT. These are explicitly given by the following Killing vectors in AdS$_3$ \cite{MaSt}:
\ba
&& L_0=i\de_+, \ \ \ti{L}_{0}=i\de_{-},  \no
&& L_{\pm 1}=ie^{\pm ix^+}\left[\f{\cosh2\rho}{\sinh2\rho}\de_{+}-\f{1}{\sinh2\rho}\de_{-}
\mp\f{i}{2}\de_\rho\right],\no
&& \ti{L}_{\pm 1}=ie^{\pm ix^-}\left[\f{\cosh2\rho}{\sinh2\rho}\de_{-}-\f{1}{\sinh2\rho}\de_{+}
\mp\f{i}{2}\de_\rho\right], \label{Lth}
\ea
where $x^\pm\equiv t\pm \phi$ and $\de_{\pm}\equiv \frac{\de}{\de x^\pm}$.

 Notice that a $SL(2,R)$ subgroup of $SL(2,R)_L\times SL(2,R)_R$ preserves the time slice $t=0$. It is generated by $l_{n}\equiv L_n-\ti{L}_{-n} (n=0,-1,1)$, i.e.
\ba
&& l_0=L_0-\ti{L}_0=i\de_\phi, \no
&& l_{-1}=L_{-1}-\ti{L}_{1}=ie^{-i\phi}\left[\f{1+\cosh(2\rho)}{\sinh(2\rho)}\de_\phi+i\de_\rho\right], \no
&& l_{1}=L_{1}-\ti{L}_{-1}=-ie^{i\phi}
\left[-\f{1+\cosh(2\rho)}{\sinh(2\rho)}\de_\phi+i\de_\rho\right].  \label{wer}
\ea
Indeed they satisfy the $SL(2,R)$ algebra and correspond to Killing vectors of the hyperbolic space $H_2$:
\be
ds^2=L^2(d\rho^2+\sinh^2\rho d\phi^2).
\ee
 We can identify the geometrical action with a linear combination of $l_n$ as follows
\ba
&& i\de_{\rho}=-\f{i}{2}(e^{i\phi}l_{-1}-e^{-i\phi}l_1),\no
&& i\de_{\phi}=l_0.  \label{actionr}
\ea

The $SL(2,R)$ transformation $g(\rho,\phi)$ which takes the origin $\rho=0$ to
a point $(\rho,\phi)$ on $H_2$ is given by
\be
g(\rho,\phi)=e^{-i\phi l_0}e^{\f{\rho}{2}(l_{1}-l_{-1})}.
\ee

\subsection{Hyperbolic/De Sitter Slices in AdS$_3$} \label{dss}

If we parameterize the hypersurface (\ref{surface}) as follows
\ba
&& X_0=L\sinh\tau\sinh\eta, \no
&& X_3=L\cosh\eta,\no
&& X_1=L\cosh\tau\sinh\eta\sin\phi, \no
&& X_2=L\cosh\tau\sinh\eta\cos\phi,
\ea
then we find the metric
\be
ds^2=L^2\left(d\eta^2+\sinh^2\eta(-d\tau^2+\cosh^2\tau d\phi^2)\right).  \label{dspat}
\ee
This shows that a constant $\eta$ slice is a two dimensional de Sitter spacetime, which is accommodated in the interval $-\pi<t<0$.
If we take the $\eta=0$ limit, it becomes a light cone. To go beyond this, we can introduce the following coordinate
\ba
&& X_0=L\cosh\tau\sin\eta, \no
&& X_3=L\cos\eta,\no
&& X_1=L\sinh\tau\sin\eta\sin\phi, \no
&& X_2=L\sinh\tau\sin\eta\cos\phi,
\ea
which leads to
\be
ds^2=L^2\left(-d\eta^2+\sin^2\eta(d\tau^2+\cosh^2\tau d\phi^2)\right). \label{hypat}
\ee
This describes the hyperbolic slices of AdS$_3$.

If we consider the Euclidean AdS$_3$ (=H$_3$) defined by
\be
X^2_0=X^2_1+X^2_2+X^2_3+L^2,
\ee
we can set
\ba
&& X_0=L\cosh\tau\cosh\eta, \no
&& X_3=L\sinh\eta,\no
&& X_1=L\sinh\tau\cosh\eta\sin\phi, \no
&& X_2=L\sinh\tau\cosh\eta\cos\phi,
\ea
to reach the metric
\be
ds^2=L^2(d\eta^2+\cosh^2\eta(d\tau^2+\sinh^2\tau d\phi^2)). \label{hsl}
\ee
This describes a hyperbolic slice of Euclidean AdS$_3$.

\section{AdS/CFT from Euclidean Path-integral}

In this section, we study Euclidean path-integral description of ground state wave functions of 2d CFTs. We introduce UV cut off efficiently in a conformal invariant way and and show that we can deform the path-integral into that on a hyperbolic space H$_2$. We will argue that this hyperbolic space corresponds to a time slice of AdS$_3$.

\subsection{Euclidean Path-integral with UV cut off}

Consider a 2d CFT on R$^2$. We define the coordinate of R$^2$ to be $(z,x)$, where $z$ is the Euclidean time. We simply express all fields in the CFT as $\phi(z,x)$.
The ground state wave function $\Psi[\phi(x)]$, which is not normalized, is written as an Euclidean path-integral:
\be
\Psi[\phi(x)]=\int \prod_{z_0<z<\infty} D\phi(z,x) \cdot \delta (\phi(z_0,x)=\phi(x))\cdot  e^{-S_{CFT}(\phi)}.
\ee

To make an explicit analysis, let us consider a free scalar field theory as a toy example:
\be
S_{CFT}=\int dx dz {\cal L}_{CFT}=\int dx dz \left[(\de_z \phi)^2+(\de_x \phi)^2\right]. \label{msd}
\ee
With the boundary condition $\phi(0,x)=\phi(x)$ we can solve the equation of motion
$(\de_x^2+\de_z^2)\phi=0$ as follows:
\be
\phi(z,x)=\int^\infty_{-\infty} dk \phi(k)e^{ikx-|k|z}, \label{wfd}
\ee
where $\phi(k)$ is the Fourier transformation of $\phi(x)$. Note that here we assumed that there is no singular behavior in the limit $z\to \infty$.

The on-shell action is evaluated as
\ba
S_{on-shell}&=&4\pi \int^\infty_{0} dz \int^\infty_{-\infty}dk |k|^2 e^{-2|k|z}\phi(k)\phi(-k) \no
&=&2\pi \int^\infty_{-\infty}dk |k|\phi(k)\phi(-k). \label{paths}
\ea
Since the path-integral of quantum fluctuation only gives an overall constant factor (see e.g.\cite{Pol}), the wave function is evaluated as
\be
\Psi[\phi(x)]\propto e^{-S_{on-shell}}=e^{-2\pi \int^\infty_{-\infty}dk |k| \phi(k)\phi(-k)},
\label{grdf}
\ee
reproducing the well-known result.

\begin{figure}
  \centering
  \includegraphics[width=8cm]{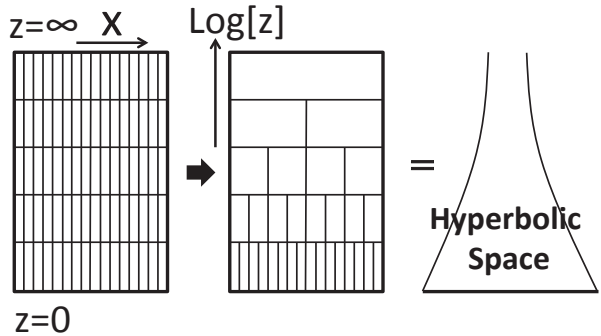}
  \caption{A computation of ground state wave function from Euclidean path-integral and its optimization, which is described by a hyperbolic geometry.}
\label{pathfig}
  \end{figure}

In this analysis we observe an important fact that
in the $k$ integral at fixed $z$ in (\ref{paths}), only modes with $|k|\lesssim 1/z$ contribute.
This fact allows us to approximate the path-integral by introducing $z$ dependent cut off of the momentum without changing the final result of wave function as depicted in Fig.\ref{pathfig}. This is realized by putting a cut off function $\Gamma(\lambda|k|z)$ in the path-integral. $\lambda$ is a parameter which controls our approximation.  We simply define it such that $\Gamma(x)=1$ if $|x|<1$, otherwise $\Gamma(x)=0$, though in our argument, the precise form of the cut off function $\Gamma(x)$ does not play an important role. The resulting wave function $\Psi[\phi(x)]$
remains approximately the same even if we introduce this $z$ dependent cut off as we will confirm below.

Let us introduce a length scale ($z_0$) dependent wave function $\Psi_{z_0}[\phi(x)]$ by the Euclidean path-integral for the range $z_0\leq z<\infty$ in the presence of the cut off $\Gamma(\lambda|k|z)$. Since $\phi(x)$ is defined by the value $\phi(z_0,x)$, we need to rescale the scalar field as $\phi\to e^{|k|z_0}\phi$ as clear from
(\ref{wfd}). Finally we obtain
\be
\Psi_{z_0}[\phi(x)]\propto e^{-2\pi \int^{\f{1}{\lambda z_0}}_{-\f{1}{\lambda z_0}}dk |k|\left(1-e^{2|k|z_0-2/\lambda}\right)\phi(k)\phi(-k)}.
\label{wfe}
\ee
First of all, it is obvious that the function (\ref{wfe}) at $z_0=0$ coincides with the correct vacuum wave function (\ref{grdf}) assuming $\lambda\ll 1$. In principle, we can also multiply the factor $(1-e^{-2/\lambda})^{-1}$
on the cut off function so that we get the correct wave function at $z_0=0$ even when $\lambda$ is $O(1)$. Having this in mind we simply set $\lambda=1$ below.

 Alternatively, we can improve this procedure by taking into account contributions from high momentum modes in a non-local way. Let us, for example, replace the cut off function as follows:
\be
\Gamma(|k|z)\to f(|k|z)\equiv \Gamma(|k|z)+\frac{1}{2}\left(1-\Gamma(|k|z)\right)\cdot e^{|k|z-1},
\label{nonl}
\ee
where we chose this such that the higher momentum modes $|k|z_0\gg 1$ are suppressed in the path-integral.

As we analyze in detail in the appendix A (as the example (ii)), in this case, the high momentum contribution cancels the extra term $\sim e^{2|k|z_0-2/\lambda}$ in (\ref{wfe}) when $|k|z_0<1$.
Finally the wave function at length scale $z_0$ reads
\be
\Psi_{z_0}[\phi(x)]\propto e^{-2\pi \int_{|k|\leq 1/z_0}dk |k| \phi(k)\phi(-k)}\times e^{
-2\pi\int_{|k|> 1/z_0}dk |k|e^{|k|z_0-1} \phi(k)\phi(-k)}.
\label{newf}
\ee
Note that for the modes below the cut off, this reproduces the correct vacuum wave function
(\ref{grdf}). On the other hand, it is clear that the higher momentum modes $|k|z_0\gg 1$
are exponentially suppressed. Thus this wave function (\ref{newf}) possesses the desired property.

In summary, for a finite value of $z_0$, this wave function (\ref{wfe}) or (\ref{newf}) describes the ground state below the cut off. On the other hand, for much higher momentum modes, it becomes trivial and thus it describes a state without any real space entanglement (i.e. boundary state as will explain in the next section)\footnote{The wave function (\ref{wfe}) corresponds to the boundary state with Neumann boundary condition, while the wave function (\ref{newf}) describes the
 Dirichlet one. For more details, refer to the appendix A.}. This nicely describes the effective wave function at the length scale $z_0$ under a real space renormalization group flow.\footnote{
As we will argue soon later in subsection \ref{locala}, in a 2d holographic CFT with a large central charge $c$, the actual momentum cut off at the length scale $z_0$ is estimated to be $|k|\lesssim\frac{c}{z_0}$ (instead of $|k|\lesssim\frac{1}{z_0}$). Therefore the momentum region $\frac{1}{z_0}\lesssim|k|\lesssim\frac{c}{z_0}$
is physically meaningful and is described by a non-local field theory.}\\

If we imagine various ways to discretize our Euclidean path-integral, this $z$-dependent UV cut off provides us with an efficient choice for this procedure. We call this an optimized Euclidean path-integral below. If we consider $K$ sites in the UV theory (corresponding to $z=\ep$), we will have $\sim K\cdot (\ep/z)$ sites as $z$ grows. Therefore we can associate the following metric of $H_2$
\be
ds^2=\f{dz^2+dx^2}{z^2}. \label{smetric}
\ee
to the Euclidean path-integral with the UV cut off. This metric is defined such that the area measured by the metric gives the number of discretized sites. The metric (\ref{smetric}) in $x$ direction is obvious from the cut off function $\Gamma(|k|z)$.
The metric in $z$ direction can be fixed by requiring that the vacuum state is invariant under the $SL(2,R)$ conformal symmetry generated by $l_0,l_{\pm 1}$. Note that this symmetry action preserves the boundary $z=0$, where we define the wave function. In other words, the discretized lattice which corresponds to our Euclidean path-integral is invariance under this $SL(2,R)$ transformation.

The cut off function can also be made manifestly conformally invariant by replacing $\Gamma(|k|z)$ with $\Gamma (z\cdot k_{H_2})$, where $k_{H_2}=\s{k_x^2+k_z^2}$ is the magnitude of the wave vector of the field configuration at each point of $(z,x)$. This cut off $\Gamma (z\cdot k_{H_2})$ is interpreted as the discretization of $(z,x)$ space such that each cell has the same infinitesimal area following the hyperbolic metric (\ref{smetric}). For example, we can take an action for our optimized Euclidean path-integral to be schematically as follows:
\be
S_{opt}=\int dxdz \Gamma(z\cdot k_{H_2}){\cal L}_{CFT},  \label{optact}
\ee
where the cut off function $\Gamma(z\cdot k_{H_2})$ is acted on all fields in the CFT.
In the next subsection, we will argue that the hyperbolic space (\ref{smetric}) corresponds to
a time slice of Euclidean AdS$_3$ as a consequence of AdS/CFT.

Next we compactify the spacial coordinate $x$ and express it as $\phi$ with the periodicity $\phi\sim\phi+2\pi$. The ground state wave function is described by the Euclidean path-integral
on the upper half of the infinite cylinder $0<z<\infty$. We can introduce a $z$ dependent UV cut off as before without chaining the final wave function. This leads to a discretized path-integral on the Poincare disk with the metric:
\be
ds^2= \f{4d\zeta d\bar{\zeta}}{(1-|\zeta|^2)^2}, \label{poid}
\ee
where
\be
\zeta=e^{-z+i\phi}.
\ee
This metric is invariant under $SL(2,R)$ transformation $l_n=L_n-\ti{L}_{-n} \ (n=0,\pm 1)$
with
\be
L_{n}=-\zeta^{n+1}\f{\de}{\de\zeta},\ \ \ \ \ti{L}_{n}=-\bar{\zeta}^{n+1}\f{\de}{\de\bar{\zeta}}.
\ee
Again we observe that the metric (\ref{poid}) agrees (up to an overall factor) with that of the time slice of Euclidean global AdS$_3$ (refer to (\ref{gads})) with the identification
$\sinh\rho=\f{2|\zeta|}{1-|\zeta|^2}$. Moreover, the  $SL(2,R)$ generators $l_0,l_{\pm 1}$
coincides with (\ref{wer}).

\subsection{Interpretations in terms of AdS/CFT}\label{locala}

As we already noted, the geometries of optimized Euclidean path-integrals reproduce the time slice of Euclidean AdS as in (\ref{smetric}) and (\ref{poid}). More precisely their metrics differ by a factor $L^2$, which is $\sim c^2$ in the Planck scale unit, from the actual AdS metric. A heuristic explanation is as follows (see also section 6). Holographic CFTs are characterized by the large central charge
$c$ and a large spectrum gap \cite{Hart}. Here we would like to turn to a simple tractable model which captures these properties: a symmetric product CFT with a large central charge $c$ on a cylinder. In such a model, the long string sector, which dominates the microstate degeneracy, can be effectively described by a CFT with a $O(1)$ central charge on a cylinder with the extended radius $O(c)$ \cite{MaSu}. Thus this gives a very fine-grained momentum quantization $P\sim n/c\ \ (n\in \mbox{Z})$ instead of $P\sim n$. In the real space, this means that the lattice spacing is smaller by a factor $\sim c$. Therefore, the actual numbers of lattice sites per unit area in the metric (\ref{smetric}) and (\ref{poid}) are $\sim c^2$. This reproduce the correct AdS$_3$ metric
up to $O(1)$ constant.

In the Euclidean AdS$_3$, we can actually find infinitely many other hyperbolic slices parameterized by $\eta$ in (\ref{hsl}). Since each of them has a different radius (given by $L\cosh\eta$), it is natural to interpret that they corresponds to different choices of the
cut off function:
\be
\Gamma\left( \f{z\cdot k_{H_2}}{\cosh\eta}\right). \label{hypcut}
\ee

In the coordinate system (\ref{hsl}), we can regard $\eta$ as an extra dimension and the AdS boundary as $\eta=\pm \infty$. In this interpretation, following a standard understanding of AdS/CFT (refer to
\cite{KR,AMR} for studies of gravity dual of CFTs on AdS spaces), the evolution of $\eta$ can be regarded as a RG flow such that the momentum scale is given by $\cosh\eta$. This consideration also justifies the action (\ref{optact}) at $\eta=0$.

Let us study the field theory more carefully by using the improved UV cut off $f(|k|z)$ given by (\ref{nonl}). Our argument in the above shows that in a 2d holographic CFT with a large central charge $c$, the actual momentum cut off at the length scale $z$ is estimated to be $|k|\lesssim\frac{c}{z}$, instead of $|k|\lesssim\frac{1}{z}$. Therefore for the momentum region $\frac{1}{z}\lesssim |k|\lesssim\frac{c}{z}$, we expect a very non-local theory whose action for the free scalar is given by $S=\int dxdz \phi\cdot f(|k_{H_2}|z)\cdot \phi$, or more generally by (\ref{optact}) with $\Gamma(|k|z)$ replaced with $f(|k|z)$. On the other hand, we can physically ignore the existence of modes above this strict UV cut off $|k|\gtrsim\frac{c}{z}$.

This non-locality occurs when we consider a small structure such that $z\gtrsim \Delta x$. For example, consider the entanglement entropy $S_A$ at a fixed (Euclidean time) $z$ with respect to a subsystem $A$ given by an interval with the length $\Delta x$. If
$\Delta x \ll z$, then we expect the volume law due to the non-locality (see \cite{LT,ST} for an explicit example in such a non-local scalar field theory):
\be
S_A\sim c\f{\Delta x}{z}, \label{sav}
\ee
where we used the fact that the effective lattice spacing is estimated as $z/c$.
This agrees with the holographic computation in the Poincare AdS$_3$ as the minimal surface
almost coincides with the original interval $A$ if $\Delta x \ll z$. This is an additional support for our argument. Refer also to Fig.\ref{merafig} for another explanation of (\ref{sav}) by using a MERA tensor network.

Finally we would like to emphasize again that the above argument can be applied only to large $c$ CFTs.
For example, if we consider a CFT with a central charge $O(1)$, the strict UV cut off is given by
$|k|\leq 1/z$ and thus we cannot reach the volume law phase (\ref{sav}).

\subsection{Einstein-Rosen Bridge from Path-Integral}

Another interesting example will be the quantum state in the thermo-field double CFT given by
\be
|\Psi_{TFD}\lb\propto \sum_{n}e^{-\frac{\beta}{4}(H_1+H_2)}|n\lb_1|n\lb_2. \label{tfd}
\ee
In the Euclidean path-integral formalism, the wave function for this state is described by a path-integral on a cylinder with a finite width: $-\f{\beta}{4}<z<\f{\beta}{4}$ in the Euclidean time direction (we will set the UV cut off $z_0$ to zero for simplicity). We express the spacial coordinate as $x$.

Again we consider a free scalar field theory as a toy model for our explanation. We introduce the boundary conditions for the two boundaries at $z=\pm\f{\beta}{4}$ for the field $\phi(z,x)$:
\be
\phi\left(-\f{\beta}{4},x\right)=\phi_1(x), \ \ \ \  \phi\left(\f{\beta}{4},x\right)=\phi_2(x).
\ee
The classical solution to the equation motion $(\de_x^2+\de_z^2)\phi=0$ is given by
\be
\phi(x,z)=\int^\infty_{-\infty} dk \left[\phi_{+}(k)e^{ikx}\f{\cosh(|k|z)}
{\cosh\left(|k|\beta/4\right)}-\phi_{-}(k)e^{ikx}\f{\sinh(|k|z)}
{\sinh\left(|k|\beta/4\right)}\right],
\ee
where $\phi_{\pm}(k)$ is the Fourier transformation of $\f{1}{2}(\phi_{1}(x)\pm \phi_2(x))$.

From this expression, we can estimate the effective momentum cut off as
\be
|k|\lesssim\mbox{max}\left\{\f{1}{|z+\beta/4|},\f{1}{|z-\beta/4|}\right\}.
\ee
Therefore it is clear that near the two boundaries $z=\pm\f{\beta}{4}$, the metric of the discretized
path-integral behaves like $ds^2\propto \f{dz^2+dx^2}{(z\pm\beta/4)^2}$. The space gets maximally squeezed at the middle $z=0$. To find the precise metric, we require the $SL(2,R)$ conformal symmetry, which leads to the metric
\be
ds^2=d\rho^2+\cosh^2\rho d\phi^2, \label{ERB}
\ee
with the coordinate transformation
\be
\tan\left(\f{\pi z}{\beta}\right)=\tanh\left(\f{\rho}{2}\right).
\ee
The SL(2,R) symmetry is explicitly given by
\ba
&& l_0=\de_\phi, \no
&& l_{-1}=e^\phi\left(\f{\cosh2\rho -1}{\sinh2\rho}\de_\phi-\de_\rho\right), \no
&& l_{1}=e^{-\phi}\left(\f{\cosh2\rho -1}{\sinh2\rho}\de_\phi+\de_\rho\right).
\ea
The metric (\ref{ERB}) coincides with the time slice of BTZ black hole i.e. the Einstein-Rosen bridge,
which is known to be dual to the thermo-field double state (\ref{tfd}) \cite{MaE}.

\subsection{More General Backgrounds}

Now we would like to turn to an Euclidean path-integral description of more general states.
First, let us introduce a mass gap. For example, consider adding the mass term
$m^2\phi^2$ in (\ref{msd}). Then the factor $e^{-|k|z}$ in (\ref{wfd}) is replaced with
$e^{-\s{k^2+m^2}z}$. This shows that we can ignore the path-integral for the region
$z\gg 1/m$. Thus for an optimized lattice computation, the region $z\gg 1/m$ should removed.
This is qualitatively consistent with time slices of holographic geometries dual to confining gauge theories.

Next consider a static excited state in a CFT such as the primary state. We can insert a primary field at $z=\infty$ and perform the path-integral until we reach $z=0$.
As in previous sections, we can introduce $z$ dependent UV cut off without changing the final wave function, which makes discretized computations efficient. Since in such an example, the $SL(2,R)$ conformal symmetry is broken and thus quantitative analysis is not straightforward, we would like to briefly discuss only qualitative aspects here. In general, to describe excited states we need more discretized lattices and the associated metric is increased compared with that for the vacuum. This is because to describe an insertion of operator at $z=z_1$ with a high momentum scale such that $k_1z_1\gg 1$, the original UV cut off $\Gamma(|k|z)$ is not enough and should be fined grained. To describe deconfined states ($\Delta>\f{c}{24}$), we need
large number $\sim 2\pi\s{c\Delta/6}$ of lattice sites even at $z=\infty$. Therefore in this case $z=\infty$ is interpreted as a black hole horizon.

More generally, if we consider time dependent excited states, it gets more difficult to find a definite connection between the optimized Euclidean path-integral and its gravity dual, mainly because there is no criterion how to choose nice time-slices. Nevertheless, a natural generalization of our previous argument is that for each of all time slices, we can associate an Euclidean path-integral with a position dependent UV cut off.

\subsection{Lorentzian AdS/CFT}

So far we have discussed an interpretation of AdS/CFT for the Euclidean AdS$_3$.
As a next step we would like to move on to the Lorentzian AdS$_3$, which will be the main focus in the rest of this paper. In Lorentzian AdS/CFT, we expect that each codimension two surface corresponds
to a quantum state with the unit norm as argued in \cite{MT}, called surface/state correspondence. Therefore, in AdS$_3$, the evolution of a closed curve on a time slice correspond to that of quantum states and we expect that this is described by a continuous version of tensor network. Indeed, as shown in \cite{TNR,TNRMERA}, a procedure called tensor network renormalization tells us that the Euclidean path-integral can be (up to overall normalization) well approximated by MERA network \cite{MERA}.
It is intriguing to note that this tensor network renormalization looks very analogous to our procedure of Euclidean path-integral summarized in Fig.\ref{pathfig}, though the former is formulated in the language of tensor networks.

Since the (minimum) time slice in the Euclidean AdS$_3$ is the same as that in the Lorentzian one, we expect that the time slice in Lorentzian AdS$_3$ correspond to a tensor network for the CFT vacuum which can be approximated by MERA. We can apply the same argument for time-independent excited states in holographic CFTs (i.e. primary states and their descendants).

Consider an analytical continuation from the Euclidean AdS$_3$ to Lorentzian one.
If we perform the Wick rotation $(\tau,\eta)\to (\tau,i\eta+i\pi/2)$, then the Euclidean AdS$_3$ metric (\ref{hsl}) is mapped to the Lorentzian one with the hyperbolic slices (\ref{hypat}), while the shift
$(\tau,\eta)\to (\tau+i\pi/2,\eta+i\pi/2)$ maps (\ref{hsl}) into the de Sitter slices (\ref{dspat}).
If we take an analytical continuation of the UV cut off function (\ref{hypcut}) on the hyperbolic slice in Euclidean AdS$_3$, we find for the Lorentzian AdS$_3$:
\ba
\mbox{Hyperbolic Slices}: \Gamma\left(\frac{k_{H_2}\cdot z}{\sin\eta}\right), \ \ \ \
\mbox{De Sitter Slices}: \Gamma\left(\frac{k_{dS_2}\cdot z}{\sinh\eta}\right). \label{cutsl}
\ea

In order to have a direct contact with explicit field theories, we would like to work with a continuum version of tensor network with UV cut off, rather than explicit lattice models. Therefore we would like to study a continuous description of tensor network, especially focusing on cMERA (continuous MERA) \cite{cMERA}. Indeed in the next section we will see that for the free scalar example, the length scale dependent wave function (\ref{wfe}) is essentially the same as that in a cMERA  description.

\section{cMERA}

In this section we review the formulation of cMERA for CFTs \cite{cMERA}
with several elaborations e.g. constructions of cMERA for a massless scalar field theory on a circle
and details of the space-like scaling operation, aiming at an interpretation of Lorentzian AdS/CFT.

\subsection{Construction of cMERA}

The formulation of cMERA is originally introduced in \cite{cMERA} for field theories on
the non-compact spacetime R$^{d+1}$. We start from the IR state $|\Psi_{IR}\lb$ which is
completely disentangled so that its real space entanglement is vanishing below the UV cut off scale. To fix our convention we write the lattice constant as $\ep$ which is the inverse of the UV cut off $\Lambda=1/\ep$. As in MERA \cite{MERA}, we add the entanglement at each length scale
and perform the coarse-graining until we finish this procedure at the UV cut off scale.

We specify the length scale by $u$ such that this corresponds to the coarse-graining by the factor $e^{u}$. This parameter $u$ takes the values from $u=-\infty$ (IR limit) to $u=0$ (UV limit).
A spacial region with the linear size $y$ in the UV theory ($u=0$) is regarded as that with the linear size $ye^u$ at the scale $u$. Note that we keep the UV cut off $\ep$ unchanged. Therefore the number of lattice sites gets decreased as $e^{du}$ when we goes from UV to IR. Therefore when the space manifold on which our field theory is defined, is compact, the lattice points get trivialized in the IR limit. Therefore we can regard $|\Psi_{IR}\lb$ as a disentangled state with no real space entanglement. For CFTs on non-compact spacetimes such as $R^{d+1}$, there is a subtlety to define
$|\Psi_{IR}\lb$ as the number of lattice sites is still infinite even in the IR limit. Therefore in this case $|\Psi_{IR}\lb$ turns out to be equal to the CFT vacuum $|0\lb$ (for the modes below UV cut off scale, i.e. $k\leq \Lambda=1/\ep$). As analyzed in \cite{cMERA} if we introduce a massive deformation, this subtlety does not happen and $|\Psi_{IR}\lb$ is given by the disentangled state.

The operation (so called entangler) which adds entanglement is written as $K(u)$, which is an integral of an operator which is local below the UV cut off scale $\ep$. The coarse-graining operation is
described by $L$, which is space-like (non-relativistic) scale transformation as we will see later in more details. The state at scale $u$, denoted by $|\Psi(u)\lb$ is expressed as follows:
\ba
|\Psi(u)\lb=P\exp\left[-i\int^u_{-\infty} d\ti{u}(K(\ti{u})+L)\right]|\Psi_{IR}\lb.
\ea
As mentioned earlier, for CFTs on $R^{d+1}$, the IR state $|\Psi_{IR}\lb$ is equal to the vacuum
$|0\lb$.

One immediately notice that the choice of $K(u)$ has huge ambiguities if we just fix the
IR state and UV state. There are infinitely many ways to interpolate the two states. However, for a CFT vacuum state, we can choose a special one owing to the conformal symmetry. This canonical choice is such that $K(\ti{u})+L$ coincides with the dilatation operator
or equally relativistic scale transformation denoted by $L'$ for the modes below the UV cut off scale.
Explicity, we have $L'=\int dx^d \sum_{i=1}^d T_{tx_i}x^i$, using the energy stress tensor $T_{\mu\nu}$.
Above the UV cut off scale $k>\Lambda=1/\ep$, we set $K=0$, while $L$ is still present. For the details of this refer to the original paper \cite{cMERA} and further studies in \cite{VP}. In most parts of our arguments we will not write the higher modes explicitly. Thus in the cMERA construction we have $|\Psi(u)\lb$ is given by the CFT vacuum below the UV cut off scale.

It is sometimes useful to introduce `interaction picture' counterpart of the above cMERA formulation based on the state  $|\Phi(u)\lb$ \cite{NRT}, which is simply related to $|\Psi(u)\lb$ via
\be
|\Psi(u)\lb=e^{-iuL}|\Phi(u)\lb. \label{tranfg}
\ee
This state is expressed as follows:
\ba
|\Phi(u)\lb=P\exp\left[-i\int^u_{-\infty} d\ti{u} \hat{K}(\ti{u})\right]|\Phi_{IR}\lb,
\label{diskh}
\ea
where we defined $\hat{K}(u)=e^{iuL}K(u)e^{-iuL}$. In this description, the effective momentum cut off is $u$ dependent as $\Lambda e^u$, while the size of the space manifold does not change.
In this description of the compactification radius does not depend on $u$.

\subsection{Free massless scalar field theory}

Consider an example of free massless scalar field theory on $R^{d+1}$. We will rewrite the original formulation of \cite{cMERA} in terms of creation and annihilation operators as in \cite{NRT}.
The Hamiltonian of this theory is defined by
\be
H=\f{1}{2}\int dk^d \left[\pi(k)\pi(-k)+|k|^2\phi(k)\phi(-k)\right].
\ee

We can define the creation and annihilation operator of the scalar field: $a_{k}$ and $a_{k}^{\dagger}$
as follows
\ba
&& \phi(k)=\f{a_k+a^\dagger_{-k}}{\s{2|k|}}, \no
&& \pi(k)=\s{2|k|}\left(\f{a_k-a^\dagger_{-k}}{2i}\right),
\ea
so that they satisfy $[a_{k},a^{\dagger}_{k'}]=\delta^d(k-k')$.

In the description by $|\Psi(u)\lb$, the action of the operation $K+L=L'$ for the modes below the UV cut off scale (i.e. $k\leq \Lambda=1/\ep$), which is equal to the relativistic scale transformation, is defined by
\ba
&& L'=-\f{1}{2}\int dx^d \left[\pi(x)x\de_x\phi(x)+x\de_x\phi(x)\pi(x)+\f{d-1}{2}\phi(x)\pi(x)
+\f{d-1}{2}\pi(x)\phi(x)\right],
\ea
and leads to the action on the annihilation operator (similarly on the creation operator)
\ba
e^{-iuL'}a_k e^{iuL'}=e^{-\f{d}{2}u}a_{ke^{-u}}. \label{sca}
\ea
Note that in two dimensional case, which we are interested in our paper, $L'$ coinsides with the dilatation charge $\int dx T_{tx}x$ as expected.

The non-relativistic and relativistic scale transformation are defined as follows
\cite{cMERA}
\ba
&& L=-\f{1}{2}\int dx^d \left[\pi(x)x\de_x\phi(x)+x\de_x\phi(x)\pi(x)+\f{d}{2}\phi(x)\pi(x)
+\f{d}{2}\pi(x)\phi(x)\right],
\ea
and its action is given by
\ba
e^{-iuL}a_k e^{iuL}=e^{-\f{d}{2}u}(\cosh(u/2)a_{ke^{-u}}+\sinh(u/2)a^{\dagger}_{-ke^{-u}}).\label{kkw}
\ea

In the description by $|\Phi(u)\lb$, the IR state $|\Phi_{IR}\lb$ should be a state with no real space entanglement. It can be constructed as the ground state of the highly massive Hamiltonian $H_\Lambda=\f{1}{2}\int dx [\pi(x)^2+\Lambda^2 \phi(x)^2]$ because in the IR limit the Hamiltonian becomes infinitely many copies of harmonic oscillators corresponding to each lattice points. This state is invariant under the transformation by $L$ and is constructed explicitly as follows. The ground state condition is written as $a_x|\Phi_{IR}\lb=0$, where $a_x=\s{\Lambda}\phi(x)+\f{i}{\s{\Lambda}}\pi(x)$ is the annihilation operator in real space. By taking Fourier transformation, we can express it as
\be
|\Phi_{IR}\lb=\prod_{k<\Lambda}|\Omega^k_{\Lambda}\lb,
\ee
where $|\Omega^k_{\Lambda}\lb$ is defined by the condition:
\be
(\ap_k a_k+\beta_k a^{\dagger}_{-k})|\Omega^k_{\Lambda}\lb=0, \label{irs}
\ee
where
\ba
&&\ap_k=\f{1}{2}\left(\s{\f{\Lambda}{|k|}}+\s{\f{|k|}{\Lambda}}\right),\ \
\beta_k=\f{1}{2}\left(\s{\f{\Lambda}{|k|}}-\s{\f{|k|}{\Lambda}}\right).
\ea

Assuming that the state is ``Gaussian'', the entangler $\hat{K}$ (\ref{diskh}) takes the following form
\be
\hat{K}(u)=\f{i}{4}\int dk^d \Gamma(ke^{-u}/\Lambda)\left(a^{\dagger}_{k}a^{\dagger}_{-k}
-a_{k}a_{-k}\right), \label{Ksc}
\ee
$\Gamma(x)$ is the cut off function such that $\Gamma(x)=1$ when $x\leq 1$ and $\Gamma(x)=0$ for $x>1$. This operation $\hat{K}$ induces the correct Bogoliubov transformation which maps the IR disentangled state $|\Phi(-\infty)\lb=|\Phi_{IR}\lb$ into the CFT vacuum $|\Phi(0)\lb=|0\lb$. More explicitly we find
\be
|\Phi(u)\lb\propto\left[\prod_{k<\Lambda e^u}\exp\left(\tanh\f{u}{2}a^\dagger_k a^\dagger_{-k}\right)|0_k\lb\right]\cdot \left[\prod_{k>\Lambda e^u}|\Omega^k_\Lambda\lb \right].
\ee

On the other hand, $|\Psi(u)\lb$ is explicitly given by
\be
|\Psi(u)\lb=\left[\prod_{k<\Lambda}|0_k\lb\right]\cdot \left[\prod_{k>\Lambda}|\Omega^k_\Lambda\lb \right].
\label{psista}
\ee
Remember that even though the right-hand side looks $u$ independent, the definition of momentum $k$ is $u$ dependent such that the actual unit length scale between lattice sites grows like $e^{-u}$. Therefore, the UV cut off $\Lambda$ corresponds to the actual momentum scale $\Lambda e^u$. In addition, $|\Omega^k_\Lambda\lb$ represents a trivial state with no real space entanglement. Therefore we can identify the quantum state defined by the wave function $\Psi_z$ (\ref{wfe}) essentially coincides with the cMERA state $|\Psi(u)\lb$ (\ref{psista}).\footnote{To see this explicitly, after the shift of momentum $k\to ke^u$ with the standard identification $z_0=\ep\cdot e^{-u}$, the quantity inside the exponential in (\ref{wfe}) becomes $\int^{\Lambda/\lambda}_{-\Lambda/\lambda} dk |k|(1-e^{-2/\lambda+2|k|\ep})\vp(k)\vp(-k)$. For $|k|\ep\ll 1$ we have the vacuum state, while for
$|k|>\Lambda$ we have the trivial wave function corresponding to the Neumann boundary state \cite{Is,CS}. As we discuss as the example (i) in appendix A, we can improve the high momentum behavior of the cut off function and realize the state $|\Omega_\Lambda\lb$ for $|k|\gg\Lambda$. The one more other choice (ii) in the appendix A or equally (\ref{nonl}) and (\ref{newf}), leads to the Dirichlet boundary state $|B_D\lb$ for $|k|\gg\Lambda$, which is similar to a version of cMERA considered in \cite{MRTW}: $|\Psi(u)\lb=\left[\prod_{k<\Lambda}|0_k\lb\right]\cdot \left[\prod_{k>\Lambda}|B_D\lb \right].$} Combined with the argument in the previous section based on Euclidean path-integral, this observation strongly suggests that the hyperbolic time slice in the Lorentzian AdS corresponds to cMERA.

\subsection{Compactification and Space-like Scale Transformation}

For the purpose of this paper, it is very useful to compactify the space coordinates. For simplicity, we would like to focus on two dimensional CFTs on R$\times$S$^1$, where the space coordinate is compactified on a circle S$^1$. We take the radius of the circle in the original UV theory to be $R_0$. In the $|\Psi(u)\lb$ picture, the radius depends on $u$ as $R(u)=R_0 e^u$, while in the $|\Phi(u)\lb$ picture, the radius is independent from $u$ as $R(u)=R_0$.

Now we would like to turn to the space-like scale transformation $L$.
In general QFTs, we argue that the $L$ action is simply given by a specific quantum quench where the metric in the space direction changes:
\be
H=H_{0}+\int dx^d \sum_{i,j=1}^d\delta g_{ij}T^{ij}=H_{0}-\int dx^d \sum_{i,j=1}^d \delta g^{ij}T_{ij},
\ee
where we take
\be
\delta g_{ij}=2 \eta \delta _{ij}.
\ee
This shift of Hamiltonian changes the radius $R$ into $R(1+\eta)$.

For example, consider a free scalar field CFT $\phi(t,x)$ in two dimension.
The action looks like
\be
S=\int dt dx \left[\f{1}{2}(\de_t\phi)^2-\f{1}{2R^2}(\de_x\phi)^2\right],
\ee
and the Hamiltonian is found to be ($\pi=\dot\phi$)
\be
H=\int dx\left[\f{1}{2}\pi^2+\f{1}{2R^2}(\de_x\phi)^2\right].
\ee
We compactify the space coordinate $x$ such that $x\sim x+2\pi$. Then the radius is given by $R$.
The mode expansion of the scalar field is given by
\ba
&& \phi(x,t)=\s{R}\sum_{n\in Z}\f{1}{\s{|n|}}\left[e^{-inx-i\f{|n|}{R}t}a_n+e^{-inx+i\f{|n|}{R}t}a^\dagger_{-n}\right], \no
&& \pi(x,t)=\f{i}{\s{R}}\sum_{n\in Z}\s{|n|}\left[-e^{-inx-i\f{|n|}{R}t}a_n+e^{-inx+i\f{|n|}{R}t}a^\dagger_{-n}\right], \no
\ea
where the canonical commutation relation is given by
\be
[a_n,a^\dagger_{m}]=\delta_{n,m}.
\ee

Consider a quench process where we suddenly change the radius $R$ into $R'$ at $t=0$. If we define the creation and annihilation operator in the new theory by $b_{n}$ and $b^\dagger_n$, by matching
$\phi$ and $\pi$ at $t=0$ we find
\ba
\s{R}(a_n+a^\dagger_{-n})=\s{R'}(b_n+b^\dagger_{-n}),\no
\f{1}{\s{R}}(a_n-a^\dagger_{-n})=\f{1}{\s{R'}}(b_n-b^\dagger_{-n}).\no
\ea

 Now, the transformation $e^{-iuL}$ change the radius from $R$ to $Re^u$. Thus if we set $R'=Re^u$ we find the transformation
 \be
 a_n=\cosh\f{u}{2}\cdot b_n+\sinh\f{u}{2}\cdot b^\dagger_{-n}, \label{Ltra}
 \ee
which agrees with (\ref{kkw}).

This leads to the following transformation rule:
\ba
e^{-iuL}a_n e^{iuL}=\cosh(u/2)a_{n}+\sinh(u/2)a^{\dagger}_{-n}.\label{kkww}
\ea
Indeed this reproduces the non-compact limit result (\ref{kkw}) by setting $k=n/R$.

As we have explained before, below the $u$ dependent UV cut off
\be
n\leq R_0 e^u/\ep, \label{cuto}
\ee
we simply have $|\Psi(u)\lb=|0\lb$ in the $|\Psi(u)\lb$ picture\footnote{In the $|\Phi(u)\lb$ description, we find the constraint $(\cosh(u/2)a_{n}-\sinh(u/2)a^{\dagger}_{-n})|\Phi(u)\lb=0$.
Thus we can identify (here again we omit the higher momentum modes $n>R_0 e^u/\ep$)
\be
|\Phi(u)\lb=\exp\left(\tanh\f{u}{2}\sum_{n<R_0 e^u/\ep} a^\dagger_n a^\dagger_{-n}\right)|0\lb.
\ee
}
(or equally see (\ref{psista})). This is obvious from the $L'$ action (below the UV cut off scale):
\ba
&& e^{-iuL'}a_n e^{iuL'}=a_{n}.\no
\label{kkwq}
\ea

In this free scalar model, we can confirm\footnote{Note that the latter two identities in (\ref{lln}) only holds below the UV cut off scale.}
\be
[L,l_n]=0, \ \ \mbox{and}\ \ \  [K,l_n]=[K+L,l_n]=0, \label{lln}
\ee
where $l_n\equiv L_n-\ti{L}_{-n}$. $L_n$ and $\ti{L}_n$ are the Virasoro generators in the left and right-moving sector and they are explicitly given by
\be
L_n=\f{1}{2}\sum_{m\in Z}\ap_{m}\ap_{n-m},\ \ \ \ti{L}_n=\f{1}{2}\sum_{m\in Z}\ti{\ap}_{m}\ti{\ap}_{n-m},
\ee
where we defined $\ap_n$ and $\ti{\ap}_n$ such that when $n>0$, $i\ap_n=\s{n}a_n$ and $i\ti{\ap}_n=\s{n}a_{-n}$ and that when $n<0$, $-i\ap_n=\s{-n}a^\dagger_{-n}$ and $-i\ti{\ap}_{-n}=\s{n}a^\dagger_{n}$.

We expect that these properties are true in general cMERA for two dimensional CFTs.
In the IR limit ($u\to \infty$),  $|\Psi(u)\lb$ approaches a $L$ invariant state:
$L|\Psi(-\infty)\lb=0$. Therefore we expect from (\ref{lln}) it satisfies the following identity:
\be
l_n|\Psi(-\infty)\lb=0.
\ee
Thus $|\Psi(-\infty)\lb$ is given by a boundary state.\footnote{In the definition of (\ref{irs}),
it corresponds to the Neumann boundary condition.} Since we do not expect any excitation of the primary state, it is natural to identify it with an Ishibashi state \cite{Is} for the vacuum sector \cite{MNSTW,MRTW}.

Even if we ignore the connection to cMERA, our argument here shows the following intriguing fact: the evolution from the CFT vacuum to the Ishibashi state $|\Phi_{IR}\lb$ is realized as a quantum quench induced by the radius change.

\section{Continuous Tensor Networks and AdS$_3/$CFT$_2$}

In the paper \cite{MT}, it has been conjectured that there is a holographic map between any codimension two convex surface $\Sigma$ in a gravitational spacetime and a quantum state in a dual Hilbert space, called surface/state correspondence. This generalizes the standard holographic principle as the gravitational spacetime does not necessarily need the presence of a boundary. If we apply this duality to the AdS$_3/$CFT$_2$, we find that each convex closed curve $\Sigma$ corresponds to a quantum state
$|\Psi(\Sigma)\lb$ in the dual CFT Hilbert space. By considering a foliation by closed curves, we obtain a tensor network for each codimension one slice in AdS$_3$. We gave a general formulation of
continuous tensor in the appendix B. Clearly, a particularly simple and nice example of codimension one slice is the time slice $t=$const which we will study in detail below.

This surface/state correspondence was originally motivated by assuming a possible description of gravitational spacetimes by ideal tensor networks. Later in \cite{MNSTW}, this tensor network is argued to be described by cMERA mainly from the viewpoint $|\Phi(u)\lb$ description. Here we would like to
 study this issue of AdS$_3/$CFT$_2$ in the $|\Psi(u)\lb$ picture.

Our argument in this section goes as follows. We start with the surface/state correspondence for AdS$_3/$CFT$_2$. Using the Killing symmetry and its holographic counterpart in CFT$_2$ we will construct a continuous tensor network. Next we show that this tensor network actually coincides with that of cMERA with the canonical choice of $K$ (below the cut off scale).

\subsection{Constructing Continuous Tensor Network from AdS$_3/$CFT$_2$}

Now we would like to construct a continuous tensor network which describes the global AdS$_3$ spacetime (\ref{gads}) via the surface/state correspondence. We focus on the closed curve defined by a constant value of $\rho$ on the time slice $t=0$. We write the corresponding state parameterized by the value of $\rho$ as $|\Psi(\rho)\lb$. The dual CFT in AdS$_3/$CFT$_2$ lives on the boundary of AdS$_3$  parameterized by the boundary coordinate $(t,\phi)$ and the radius $R$ of space coordinate
$\phi$ is $R=1$. If we express the UV cut off of CFT as that of the AdS space given by
$\rho=\rho_\infty(\to\infty)$, the quantum state $|\Psi(\rho)\lb$ is defined in the Hilbert space of
dual CFT on R$\times$S$^1$ with the radius of circle S$^1$ given by
\be
R(\rho)=\frac{\sinh\rho}{\sinh\rho_\infty}.
\ee
Note that this Hilbert space is always regularized by a lattice spacing $\ep$ which does not depend on $\rho$. The UV state $|\Psi(\rho_\infty)\lb$ should coincide with the CFT vacuum $|0\lb_{R=1}$, where we make the radius explicit as a subscript. The state at general $\rho$ can be written in the following form
\be
|\Psi(\rho)\lb=P\exp\left[-i\int^\rho_{0} d\ti{\rho} M(\ti{\rho})\right]|\Psi(0)\lb. \label{mnet}
\ee

We would like to determine the $\rho$ evolution operator $M(\rho)$ by employing the $SL(2,R)$ symmety we discussed just before. First remember that
\be
l_n=L_n-\ti{L}_{-n}=-\int^{2\pi}_0 d\phi e^{in\phi}T_{--}(\phi)+\int^{2\pi}_0 d\phi e^{in\phi}T_{++}(\phi),
\ee
where $T_{\mu\nu}$ is the energy stress tensor of the two dimensional CFT.

In order to find $M(\rho)$ we would like evaluate $\de_\rho$ for the infinitesimally short interval
$\phi_0-\delta/2 \leq \phi \leq \phi_0+\delta/2$. This leads to
\ba
 M(\rho)&=&\int^{2\pi}_0 d\phi_0 M(\rho,\phi_0),  \no
 M(\rho,\phi_0)&=&\f{1}{\delta}\int^{\delta/2}_{-\delta/2}d\phi\sin(\phi)\left(-T_{--}(\phi_0+\phi)
+T_{++}(\phi_0+\phi)\right), \no
&=& \f{1}{\delta}\int^{\delta/2}_{-\delta/2}d\phi\sin(\phi)T_{t\phi}(\phi_0+\phi) \no
&\simeq & \f{1}{\delta}\int^{\delta/2}_{-\delta/2}d\phi \phi T_{t\phi}(\phi_0+\phi) \simeq D(\phi_0),
\label{mnetd}
\ea
where $D$ is the dilatation operator. This agrees with the cMERA for the Poincare AdS near the AdS
boundary as $L'$ is the relativistic scale transformation as we noted before. If we perform integration of $\phi_0$ to find $M(\Sigma_\rho)$ we simply find
\be
M(\rho)=0,   \label{kzero}
\ee
in the $\delta\to 0$ limit. This can also be found from the total derivative structure
\be
 M(\rho,\phi_0)\simeq \f{\delta^2}{12}\de_\phi T_{t\phi}|_{\phi=\phi_0},
\ee
in the $\delta\to 0$ limit. This trivial evolution (\ref{kzero}) agrees the free scalar construction on a cylinder (\ref{kkwq}). Since this property is also true in cMERA for any CFTs, we find that our continuous tensor network (cTN) which is obtained from AdS$_3/$CFT$_2$, agrees with that of cMERA with the canonical choice of entangler $K$ via the identification of radius
\be
R_0 e^u=\frac{\sinh\rho}{\sinh\rho_\infty}=R(\rho). \label{udef}
\ee
Here we need to understand that this correspondence is confirmed below the UV cut off scale ($k<\Lambda$) as we do not know how to probe quantum states from AdS$_3$ above the cut off scale ($k>\Lambda$) at present. The quantum state $|\Psi(\rho)\lb$ of the network for the CFT vacuum is simply given by
\be
|\Psi(\rho)\lb=|0\lb_{R(\rho)}, \label{vactr}
\ee
where $|0\lb_{R}$ denotes the vacuum for the CFT on a cylinder with radius $R$.

In the above argument, we consider a deformation of the AdS boundary into a smaller circle with
the rotational symmetry. More generally, we can consider any deformation without any rotational symmetry by locally acting $\de_\rho$ and $\de_\phi$ in principle
such as the action $\int^{2\pi}_0 d\phi_0 [A(\phi_0)\de_\rho+B(\phi_0)\de_\phi]$. This leads to
a state $|\Psi(\Sigma)\lb$ for any closed curve $\Sigma$, which realizes the idea of surface/state correspondence.

\subsection{Other Slices}

So far we focused on the cTN on the hyperbolic plane $H_2$ defined as the time slice $t=0$.
On the other hand, the space built from MERA network is often associated with a de Sitter space
\cite{Beny,Cz,Car,VP}. Therefore it is useful to consider what kind of cTN we can obtain from de Sitter slices.

We can find one parameter family of de Sitter slices (see (\ref{dspat})) defined by:
\be
\cosh\rho\sin t=\cosh\eta_0,  \label{dst}
\ee
where $\eta_0$ is a positive constant. This is a two dimensional de Sitter space ($dS_2$) with the metric
\be
ds^2=L^2\sinh^2\eta_0(-d\tau^2+\cosh^2\tau d\phi^2),  \label{ddst}
\ee
embedded in AdS$_3$ and it approaches to $t=0$ slice at the AdS boundary $\rho\to \infty$.
We can confirm that the actions of $l_{\pm 1}$ and $l_0$ preserve the $dS_2$ (\ref{dst}) directly.\footnote{We are very grateful to Juan Maldacena for pointing out this.}
By generalizing (\ref{actionr}) for non-zero $t$ we find
\be
-\f{i}{2}(e^{i\phi}l_{-1}-e^{-i\phi}l_1)=i(\cos t \de_\rho -\sin t \tanh\rho \de_t)=i\de_\tau.
\ee
Thus as in the previous subsection, we can identify the $\tau$ evolution in (\ref{ddst}) as the
the same operation (\ref{mnetd}). This shows that the cTN on the $dS_2$ can also be identified with
cMERA network. In this way, as far as we consider the CFT vacuum, the hyperbolic slice and de Sitter slice have equivalent properties and can be identified with cMERA network except that there
is a non-zero minimum radius $\sinh\eta_0>0$ only in the latter. However, once we consider excited states in a holographic CFT, they lead to different cTN descriptions of an identical state as we will
see in the section 7.

\section{An Argument for Sub-AdS Scale Locality}\label{locali}

So far, in our analysis of continuous tensor networks, we studied general two dimensional CFTs and did not employ any special condition for holographic CFTs. Therefore, this is not enough to explain the
sub-AdS scale locality \cite{Hat,Car}. Indeed,  in our cMERA formulation, the momentum cut off appears as follows:
\be
n\leq \f{R_0e^u}{\ep}=\f{\sinh\rho}{\ep\sinh\rho_\infty}\sim \sinh\rho,
\ee
where we employed the cut off (\ref{cuto}) and the standard identification of
the UV cut off in AdS/CFT $\ep\sim e^{-\rho_\infty}$. Therefore when $\rho$ is $O(1)$,
we cannot distinguish the different points in the AdS$_3$ spacetime, thought the distance between such points is order $O(c)$ ($c$ is the central charge of the 2d CFT). This means a locality only at the AdS radius $L$ scale.

However, a standard knowledge of AdS/CFT tells us that in holographic CFTs we can get a finer resolution of spacetime up to the Planck scale. Holographic CFTs are characterized by the large central charge
$c$ and a large spectrum gap \cite{Hart}. Here we would like to turn to a simple tractable model which captures these properties: a symmetric product CFT with a large central charge $c$ on a cylinder with the radius $R_0=1$, though this is not exactly a holographic CFT dual to a standard classical gravity, strictly speaking. Such a CFT can be expressed as $M^m/S_m$ which is a symmetric product of $m$ identical CFTs, each of which is denoted by $M$. We have $c=mc_M$, where $c_M$ is the central charge of $M$.
This theory is defined as an orbifold of $M^m$ by the symmetric group $S_m$. Its twisted sectors include a so called long string sector. If we write a primary field of the CFT $M$ as $\phi(t,x)$, the long string sector is defined by a boundary condition like $\phi_a(t,x+2\pi)=\phi_{a+1}(t,x)$, where $a=1,2,\ddd,m$ distinguishes $m$ copies of the CFT $M$.

\begin{figure}
  \centering
  \includegraphics[width=8cm]{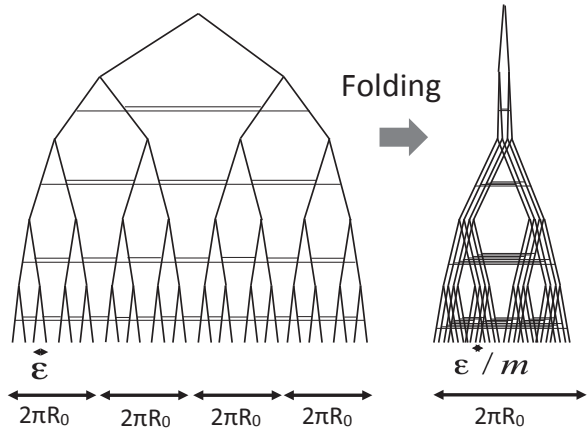}
  \caption{Folding the MERA network in a 2d symmetric product CFT $M^m/S_m$ (we chose $m=4$.). The left picture expresses a MERA network for a long string sector vacuum which is equivalent to the single string sector vacuum with the radius $mR_0=4R_0$. The right picture describes its equivalent network after the folding such that the radius is $R_0$. We show the coarse-graining (isometries) as tri-vertices and the disentanglers (unitary transformations) as horizontal lines.
  The right network shows that the actual lattice spacing is $\ep/m$.
  From this network, we can easily see that the entanglement entropy $S_A$ follows the volume law for a small interval $A$ with the width $(\ep/m\ll) \Delta x\ll \ep$ in the large $c$ limit $m\to \infty$. Note also that the final MERA network is squeezed near the top region (roughly more than $\log \frac{R_0}{\ep}$ steps from the bottom). This is very analogous to the global AdS$_3$ metric.}\label{merafig}
  \end{figure}

In this model, the long string sector, which dominates the microstate degeneracy, can be effectively\footnote{
This is clear from the boundary condition $\phi_a(t,x+2m\pi)=\phi_{a}(t,x)$.} described by a single CFT $M$ on a cylinder with the extended radius $mR_0$ \cite{MaSu}. Thus this gives a very fine-grained momentum quantization $P= \frac{n}{m}\ \ (n\in \mbox{Z})$ instead of $P=n$. After Fourier transformation, this leads to a network with a much finer structure by the factor $1/m=O(1/c)$. The resolution of this network is estimated as $L\ep \sinh\rho_\infty \cdot (1/c)\sim L/c$, which is indeed the Planck length scale. In other words, the actual lattice constant is estimated as $\ep/m$ instead of $\ep=\Lambda^{-1}$. This fact can also be explained schematically by folding a MERA network description for the long string sector as in Fig.\ref{merafig}.

The above is our heuristic argument for a sub-AdS locality in AdS$_3/$CFT$_2$. Note that in order to explain a similar locality in a higher dimensional AdS/CFT, we encounter a fractional power of central charge and this suggests that the characterization of holographic CFTs in higher dimensions is much more complicated.

\section{Bulk Locally Excited States and cMERA}

Now we would like to turn to a description of excited states in terms of continuous tensor networks. Especially we focus on a class of excited states which correspond to simple excitations in the bulk AdS, i.e. local excitations in the bulk. In the global AdS$_3$ we excite one point at a specific time.
Using the symmetry of AdS$_3$ space, we can focus on an excitation at $\rho=t=0$, constructed by acting a bulk scalar field $\vp_\ap$ on the bulk vacuum, i.e. $\vp_{\ap}|0\lb_{AdS}$. The label $\ap$ expresses the primary field in the two dimensional CFT and the bulk scalar field is also labeled by $\ap$ as $\vp_\ap$. We denote the corresponding primary state in the CFT $|\ap\lb$ and we take its chiral and anti chiral conformal dimension to be $h_\ap=\bar{h}_\ap$.

In \cite{MNSTW} (see also \cite{Ve} for a similar but different formulation), the CFT dual of such a excitation was identified with the `global Ishibashi state' with $\f{\pi}{2}$ time translation denoted by
\be
|\Psi_\ap\lb=e^{-\ti{\ep} H}e^{-i\f{\pi}{2}H}|J_\ap\lb,  \label{lest}
\ee
where $\ti{\ep}$ is the UV regularization which makes the excited energy finite and
 $|J_\ap\lb$ is the Ishibashi state for the global conformal symmetry and satisfies
\be
l_{\pm 1}|J_\ap\lb=l_0|J_\ap\lb=0.
\ee
On the other hand, the state $|\Psi_\ap\lb$ satisfies (in the limit $\ti{\ep}\to 0$)
\be
(L_{\pm 1}+\ti{L}_{\mp 1})|\Psi_\ap\lb=0,\ \ \ \ (L_0-\ti{L}_0)|\Psi_\ap\lb=0. \label{lcon}
\ee

More explicitly, $|J_\ap\lb$ is written as
\be
|J_\ap\lb=\s{\cal{N}}\sum_{k=0}^\infty e^{-\ti{\ep} k}|k,\ap\lb, \label{ishir}
\ee
where
\ba
&& |k,\ap\lb=\f{1}{N_k}(L_{-1})^k(\ti{L}_{-1})^k|\ap\lb. \no
&& N_k\equiv \f{\Gamma(k+1)\Gamma(2h_\ap+k)}{\Gamma(2h_\ap)}.  \label{ishik}
\ea
This CFT dual of bulk locally excited state is precisely identical to that obtained by acting the known CFT dual of bulk local field (HKLL map) \cite{HKLL} on the vacuum state as shown in \cite{MNSTW,NO,GMT}.

\subsection{Continuous Tensor Network for Bulk Locally Excited States}

\begin{figure}
  \centering
  \includegraphics[width=6cm]{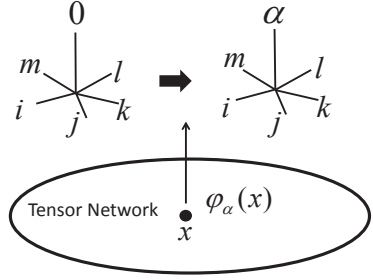}
  \caption{A modification of tensor network dual to a bulk local excitation.}
\label{tensorfig}
  \end{figure}

We would like to construct continuous tensor networks which reproduce such locally excited states in the global AdS$_3$. In order to realize the UV state $|\Psi(u)\lb$ other than the CFT vacuum state,
we obviously need to modify the tensor operator $M(\rho)$ in (\ref{mnet}). For example, in cMERA for a free scalar field such a modification is solved for quantum quench excitations in \cite{NRT}.
In terms of the discretized tensor networks for lattice quantum systems, such modification is realized by changing tensors as in \cite{Qie,HAPPY}. Especially if we excite a point in the bulk by a local field, we can obtain the tensor network by replacing a tensor located at the point where the local field is inserted as depicted in Fig.\ref{tensorfig}.

Below we would like to construct a cTN for the locally excited state following this prescription.
If we insert the bulk local field at $\rho=0$,  we can still use the same network with $M(\rho,\phi)$ given by (\ref{mnetd}) for $\rho>0$. Since we have $M(\rho)=0$ as we showed before, we can simply identify the state $|\Psi(\rho)\lb$ with
\be
|\Psi(\rho)\lb=|\Psi_\ap\lb_{R(\rho)},  \label{onet}
\ee
so that it reproduces (\ref{lest}) in the UV limit $\rho=\rho_\infty$. Note that when $|\ap\lb$ is the CFT vacuum $|0\lb$, this is trivially reduced to the network (\ref{vactr}) for the vacuum which we discussed before.

\begin{figure}
  \centering
  \includegraphics[width=6cm]{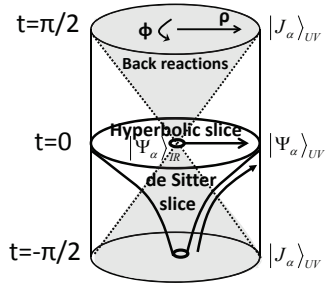}
  \caption{The tensor network evolutions which corresponds to a locally excited state in global AdS$_3$.}
\label{slicefig}
  \end{figure}

Since we focused on the hyperbolic slice $H_2$ defined by $t=0$ in the above,
one may wonder how it looks like if we choose other slices such as de Sitter slices.
In the gravity dual, if we excite a point in the bulk, the excitation expands within a light-cone as
depicted in Fig.\ref{slicefig}. Therefore we expect that the tensors which correspond to the inside light-cone region will be modified from those for the vacuum state. Since we do not have a systematic way to identify this deformation currently, we cannot closely follow cTN for general slices including de Sitter slices. In other words the hyperbolic slice has an advantage as we can only modify a single tensor (or equally the IR state) to describe the locally excited state.

Now we would like to explore the consequence of the network flow (\ref{onet}) a bit more.
Remember that the state $|\Psi_\ap\lb$ satisfies the condition (\ref{lcon}). We notice that this is
the global part of the standard property $(L_n-(-1)^n\ti{L}_{-n})|C\lb=0$
of cross cap states $|C\lb$ as noted in \cite{NO}. Moreover, if we consider the IR limit $\rho\to 0$, the allowed momentum modes are limited such that
only $n=1$ modes are meaningful as the radius $R(\rho)$ shrinks. Therefore, we can identify the IR state $\lim_{\rho\to 0}|\Psi_\ap\lb_{R(\rho)}$ as the cross cap states in holographic CFTs under a large $c$ approximation. Notice that the cross cap state is highly entangled state, while the boundary states are disentangled states \cite{MRTW}. Indeed, the latter is obtained from a time translation by $\pi/2$ of the latter state and the time evolution by $\pi/2$ leads to entanglement propagations to
anywhere as is familiar in quantum quenches \cite{GQ}.

\begin{figure}
  \centering
  \includegraphics[width=6cm]{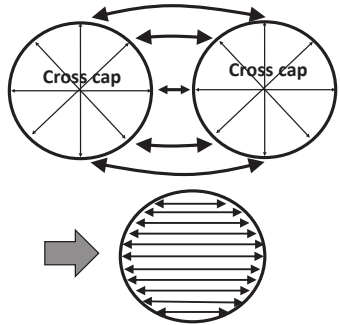}
  \caption{Gluing two cross cap states lead to an identity operation.}
\label{capfig}
  \end{figure}

In the free field CFTs, we can impose complete conditions for cross cap states by relating a point and its antipodal point in terms of fundamental free fields \cite{Is,CS}, which leads to a maximally entangled tensor if we view a state as a tensor. If we paste two cross cap states more than half of each, we get the identity operation as explained in fig.\ref{capfig}. This is the same property which the perfect tensor \cite{HAPPY} (see also \cite{Hung}) satisfies.

In the case of holographic CFTs, we only know the cross cap condition for the Virasoro generators and
explicit computations look much more difficult. However, if we remember that the state is obtained from time evolution of a boundary state, the entanglement scrambling phenomenon found in \cite{ABGH} suggests that our cross cap states in holographic CFTs are no longer such simple entangled states as in free field CFTs but are more scrambled states, similar to the random tensors in \cite{HQ}.

\section{Conclusion and Discussion}

In this paper we studied connections between tensor networks and AdS/CFT from two different viewpoints. In the first part, we considered an Euclidean path-integral description of ground state wave functions in 2d CFTs in the presence of UV cut off. We optimized the path-integral computation  by introducing a position dependent UV cut off without changing the final wave function.
 We found that this is regarded as a path-integral on a hyperbolic space and we argued that this space corresponds to a time slice of AdS. This conjecture is supported from the global symmetry of AdS$_3$ and also from the fact that such a field theory is expected to appear as a dual CFT on a hyperbolic space. By shifting its boundary, we defined a wave function as a function of effective length scale. This scale dependent wave function turns out to be essentially the same as the evolution of quantum states in cMERA at least below the cut off scale. This observation leads to the interpretation of hyperbolic time slice in (Euclidean) AdS spaces as a tensor network. It is an intriguing future problem to perform an explicit analysis of our discretized path-integral in specific lattice models.

In the latter part, we took a different approach. We started with a Lorentzian AdS$_3$ and
studied a systematic construction of continuous tensor network description dual to AdS$_3/$CFT$_2$ assuming the surface/state correspondence \cite{MT}. We obtained the resulting network for the canonical time slice, which is a hyperbolic space $H_2$, from gravitational considerations and found that it essentially coincides with the (compactification of) the original cMERA network \cite{cMERA}. However, notice that our analysis from AdS$_3$ geometry only concerns the modes below the cut off scale.

Interestingly, through an analysis of locally excited bulk state, we observed that our network is very analogous to the perfect tensor network \cite{HAPPY} and random tensor network \cite{HQ}. This is because the IR state is given by $\pi/2$ time translation of boundary state, where non-trivial quantum entanglement is generated. We also gave a heuristic argument of the sub-AdS scale bulk locality in cMERA, based on symmetric product CFTs. All these suggest that cMERA, which has full conformal invariance manifestly, can be regarded as a continuous refinement of tensor networks such as perfect and random tensor network and thus is expected to describe a canonical time slice of AdS space at least below the cut off scale.

The Wick rotation from the Euclidean AdS to the Lorentzian AdS is obvious for the canonical time slice $t=0$. The above arguments, i.e. the correspondence between an Euclidean path-integral on $H_2$ with a UV cut off and the cMERA network, are based on this fact. However, if we choose other hyperbolic slices ($\eta\neq 0$ in (\ref{hsl})) in Euclidean AdS, they can be Wick rotated into a hyperbolic (\ref{hypat}) or de Sitter (\ref{dspat}) slices. This implies that the cMERA network can also fit nicely with de Sitter slices. Indeed, our analysis based on the Killing vectors in AdS and its CFT dual support this possibility. This might give some connection to a seeming independent idea based on kinematic spaces \cite{Cz}.

In this paper we mainly consider two dimensional CFTs (2d CFTs) for simplicity. However, notice that most results can be generalized into the higher dimensional AdS/CFT in a straightforward way.

Finally it is a very interesting future problem to find an explicit relation between the spacetime metric and the property of continuous tensor network. An important challenge will be to work out how we obtain the time-like component of the metric.


\subsection*{Acknowledgements}

We would like to thank Bartek Czech, Mukund Rangamani, Frank Verastraete, Beni Yoshida for discussions and especially Glen Evenbly, Juan Maldacena and Guifre Vidal for very useful conversations.
 TT is supported by the Simons Foundation through the ``It from Qubit'' collaboration and JSPS Grant-in-Aid for Scientific Research (A) No.16H02182. MM and KW are supported by JSPS fellowshop. TT is also supported by World Premier International Research Center Initiative (WPI Initiative) from the Japan Ministry of Education, Culture, Sports, Science and Technology (MEXT).
TT is grateful to International Workshop on Tensor Networks and Quantum Many-Body Problems (TNQMP2016) held at ISSP, Tokyo U., and Integrability in Gauge and String Theory (IGST2016) held at Humboldt University, where parts of this work were presented. We would also like to thank It from Qubit activities and their organizers, especially the summer school held at Perimeter Institute and the collaboration meeting at KITP, UCSB, for stimulating discussions.

\begin{appendix}

\section{Other Choices of UV Cut Off Function}

Here we would like to consider more general (position dependent) UV cut off functions in the Euclidean path-integrals of a free scalar and compute resulting scale dependent wave functions, which generalizes our result in (\ref{wfe}). Here we set $\lambda=1$. Since we do not want to change the low momentum ($|k|z<1$) behavior we only modify the higher energy part with keeping the scaling symmetry as follows
\be
\Gamma(|k|z)\to f(|k|z)\equiv \Gamma(|k|z)+(1-\Gamma(|k|z))\cdot g(|k|z).
\ee
The wave function at the length scale $z_0$ is expressed as
\be
\Psi_{z_0}=\exp\left(-4\pi\int^{\infty}_{-\infty} dkc(k)\phi(k)\phi(-k)\right),
\ee
where
\be
c(k)=|k|^2\int^\infty_{z_0}dzf(|k|z)\cdot e^{-2|k|(z-z_0)}.
\ee
In the expression of creation/annhilation operators, this wave function is equivalent to the quantum state
\be
|\Psi_{z_0}\lb\propto \exp\left(-\int^\infty_0 dk\left(\frac{8\pi c(k)-|k|}{8\pi c(k)+|k|}\right)a^\dagger_k a^\dagger_{-k}\right)|0\lb.
\ee

When $g(|k|z)=0$, we have $c_0(k)=\frac{|k|}{2}(1-e^{2|k|z_0-2})$, which precisely corresponds to (\ref{wfe}). Below we consider two other choices.
First consider the case (i) $g(|k|z)=\frac{\beta_1}{|k|z}$, where $\beta_1$ is a constant.
In this case we obtain
\ba
c(k)&=&c_0(k)+\ap\cdot\beta_1\cdot |k|\cdot e^{2|k|z_0} \ \ \ (\mbox{if}\ \ |k|z_0<1),\no
&=& \beta_1|k|\int^\infty_0 \frac{dy}{y+z_0}e^{-2|k|y} \ \ \ (\mbox{if}\ \ |k|z_0>1).
\ea
where $\ap$ is a positive constant given by the integral $\int^\infty_1\f{dy}{y}e^{-2y}$.
When $|k|z_0\gg 1$ we can approximate
\be
c(k)\simeq \frac{\beta_1}{2z_0}.
\ee
This behavior coincides with the IR state $|\Omega_\Lambda\lb$ used in the original cMERA \cite{cMERA} (see (\ref{psista})), after the rescaling $ke^u\to k$ or equally $z_0\to \Lambda$. Thus in the connection to cMERA, this choice gives a refinement of (\ref{wfe}).

On the other hand, if we consider the second choice (ii) $g(|k|z)=\beta_2 e^{|k|z-1}$ ($\beta_2$ is an arbitrary constant), we find
\ba
c(k)&=&c_0(k)+\beta_2|k|\cdot e^{2|k|z_0-2} \ \ \ (\mbox{if}\ \ |k|z_0<1),\no
&=& \beta_2|k|\cdot e^{|k|z_0-1} \ \ \ (\mbox{if}\ \ |k|z_0>1).
\ea
In particular, we would like to choose $\beta_2=1/2$. Since in this case we get $c(k)=\frac{|k|}{2}$ for $|k|z_0<1$, this precisely reproduces the correct vacuum wave function (\ref{grdf}) in the limit $z_0=0$.
Since $c(k)$ grows exponentially, for a high momentum $|k|z_0\gg 1$ it approaches to a boundary state (Ishibashi state) for the Dirichlet boundary condition.  In the Euclidean path-integral, this choice of UV cut off can suppress high momentum modes as the scalar field action $S$ becomes very large. In this sense, it is similar to the Wilsonian renormalization group flow. These observations are consistent with the recent paper \cite{FLP}.

\section{Spacetime Continuous Tensor Network}

Here we would like to summarize a formulation of continuous tensor network (cTN), which
generalizes the cMERA formulation so that we can apply it to non-AdS spacetimes. Surface/state correspondence \cite{MT} argues that any gravitational spacetime can have such a tensor network description.
We start with a $d+2$ dimensional gravitational spacetime $M_{d+2}$ described by Einstein gravity.

We choose a coordinate $x=(t,\vec{x})\in M_{d+2}$ for simplicity, though our argument below should be independent from the choice of coordinate.
At a point $x$, we associate an entangling hermitian operator $M_i(x)$ ($i=1,2,\ddd,d+1$), which is local up to the UV cut off scale dual to the Planck length. The index $i$ describes the spacial direction of the entangling operation. The time evolution of $M_i(x)$ is simply given by a Hamiltonian $H(x)$ at each point which satisfies
\be
\f{dM_i(x)}{dt}=i[H(x),M_i(x)].
\ee

Now we take a time slice (codim. one space-like surface) $N_{d+1}$ and consider its
one parameter foliation by codim. two surfaces $\Sigma_{u}$ i.e. $N_{d+1}=\cup_u \Sigma_u$.
The surface/state duality tells us that there is a corresponding state for each surface:
\be
\Sigma_u \lr |\Psi(\Sigma_u)\lb.
\ee
Here we assume that the homology $\Sigma_{u}$ is trivial so that it is dual to a pure state.

We can write the $u$ evolution as
\be
|\Psi(\Sigma_{u_1})\lb=P\cdot \exp\left(-i\int^{u_1}_{u_2}du M(\Sigma_u)\right)|\Psi(\Sigma_{u_2})\lb.
\ee
Our basic claim is that $K(\Sigma_u)$ is expressed in terms of the local entangler $M_i(x)$ as follows
\be
M(\Sigma_u)=\int_{x\in \Sigma_u} dx^d n^i_u (x)M_i(x), \label{sum}
\ee
where $n^i_u$ is the displacement vector of $\Sigma_u$ when we change $u$ at $x$.

From (\ref{sum}) we obtain the important consistency condition our spacetime tensor network:
\be
\f{dH(x)}{du}=i[n^i_u(x)M_i(x),H(x)].
\ee

Consider various choice of foliations of the time slice $N_{d+1}$. We take two of them, expressed as $\Sigma_u$ and $\Sigma_w$ such that $N_{d+1}=\cup_u \Sigma_u=\cup_w \Sigma_w$. Let us assume that
$\Sigma_{u_1}=\Sigma_{w_1}\equiv \Sigma_1$ and $\Sigma_{u_2}=\Sigma_{w_2}\equiv \Sigma_2$.

Correspondingly we have two expressions
\ba
|\Psi(\Sigma_1)\lb &=& P\cdot \exp\left(-i\int^{u_1}_{u_2}du M(\Sigma_u)\right)|\Psi(\Sigma_2)\lb, \no
&=& P\cdot \exp\left(-i\int^{w_1}_{w_2}dw M(\Sigma_w)\right)|\Psi(\Sigma_2)\lb.
\ea
 Since we expect the same thing is also true for excited states which are obtained by replacing some of the tensor inside $\Sigma_2(\in \Sigma_1)$ locally, we require
\be
P\cdot \exp\left(-i\int^{u_1}_{u_2}du M(\Sigma_u)\right)=
P\cdot \exp\left(-i\int^{w_1}_{w_2}dw  M(\Sigma_w)\right).  \label{flatc}
\ee
 From this, we can conclude that $M(\Sigma)$ is a flat connection in the space of codim. two surfaces in $M_{d+2}$. Therefore we can write $K$ as
\be
M(\Sigma_u)=i\de_u G(u)\cdot G(u),
\ee
for a certain unitary matrix valued function $G(u)$. Then the state dual to the surface is written in the form:
\be
|\Psi(\Sigma_u)\lb=G(u)|\Psi^{(0)}\lb,
\ee
where $|\Psi^{(0)}\lb$ is a certain reference state. $G(u)$ for any condim. two surfaces in $M_{d+2}$ defines the continuous tensor network.

\end{appendix}


\end{document}